\documentclass[prl,twocolumn,preprintnumbers,amsmath,amssymb,showpacs]{revtex4}
\usepackage{graphicx, epsf,bm}
\usepackage{subfigure,amssymb,color}

\begin{document}
\title{\bf Anderson localization and the quantum phase diagram of three dimensional disordered Dirac semimetals}
\author{J. H. Pixley}
\affiliation{Condensed Matter Theory Center and Joint Quantum Institute, Department of Physics, University of Maryland, College Park, Maryland 20742- 4111 USA}
\author{Pallab Goswami}
\affiliation{Condensed Matter Theory Center and Joint Quantum Institute, Department of Physics, University of Maryland, College Park, Maryland 20742- 4111 USA}
\author{S. Das Sarma}
\affiliation{Condensed Matter Theory Center and Joint Quantum Institute, Department of Physics, University of Maryland, College Park, Maryland 20742- 4111 USA}

\date{\today}
\begin{abstract}
We study the quantum phase diagram of a three dimensional non-interacting Dirac semimetal in the presence of either quenched axial or scalar potential disorder, by calculating the average and the typical density of states as well as the inverse participation ratio using numerically exact methods. We show that as a function of the disorder strength a half-filled (i.e. undoped) Dirac semimetal displays three distinct ground states, namely an incompressible semimetal, a compressible diffusive metal, and a localized Anderson insulator, in stark contrast to a conventional dirty metal that only supports the latter two phases. We establish the existence of 
two distinct quantum critical points, which respectively govern the semimetal-metal and the metal-insulator quantum phase transitions and also reveal their underlying multifractal nature. Away from half-filling the (doped) system behaves as a diffusive metal that can undergo Anderson localization only, which is shown by determining the mobility edge and the phase diagram in terms of energy and disorder.
\end{abstract}

\pacs{71.10.Hf,72.80.Ey,73.43.Nq,72.15.Rn}

\maketitle
\maketitle
The development of Lorentz invariant relativistic quantum mechanics naturally led to the discovery of the celebrated Dirac equation~\cite{Dirac-1928} to describe the electron and positron. In high energy physics there are no massless Dirac fermions, as both leptons and quarks are massive particles and it is only at energy scales higher than the corresponding mass gap, that they can be approximated as massless.
In contrast, many condensed matter systems support a gapless phase, where two-fold Kramers degenerate conduction and valence bands touch linearly at isolated points in the Brillouin zone. The low energy excitations around these diabolic points (known as Dirac points, where the conical conduction and valence bands touch each other) can be described by a massless Dirac equation in the infrared limit, such a phase is known as a Dirac semimetal (DSM) if the system is undoped with the Fermi level precisely at the Dirac point. 

Recently, an intense experimental investigation into narrow gap semiconductors~\cite{Dornhaus-1983} has led to the discovery of three dimensional DSMs in various materials such as Cd$_3$As$_2$~\cite{Neupane-2014,Liu-2014,Borisenko-2014}, Na$_3$Bi~\cite{Liu-2014,Xu-2015}, Bi$_{1-x}$Sb$_x$~\cite{Lenoir-1996,Ghosal-2007,Teo-2008}, BiTl(S$_{1-\delta}$Se$_{\delta})_2$~\cite{Xu-2011,Sato-2011}, (Bi$_{1-x}$In$_x$)$_2$Se$_3$~\cite{Brahlek-2012,Wu-2013}, and Pb$_{1-x}$Sn$_x$Te~\cite{Dornhaus-1983,Xu-2012,Gibson-2014}. These experiments have now raised the exciting prospect of studying the physics of three dimensional massless Dirac fermions in solid state systems. Due to the ubiquitous presence of disorder in all solid state systems, a precise understanding of the phase diagram of dirty DSMs is therefore a problem of deep fundamental importance, which can only be studied in condensed matter systems since the corresponding relativistic Dirac problem in particle physics does not, by definition, have any disorder. While effects of disorder in conventional metals 
are fairly well understood~\cite{Lee-1985,Belitz-1994,Janssen-1998,Evers-2008}, this is not the case for three dimensional DSMs where the applicable quantum phase diagram and (even) the question of how many allowed phases may exist as a function of disorder are still wide open.

Due to the finite density of states (DOS) at the Fermi energy, 
a conventional metal (CM) (or a Fermi liquid) is a compressible state of matter, and an infinitesimally weak disorder acts as a relevant perturbation for the ballistic Fermi liquid fixed point, 
giving
rise to a finite lifetime for the ballistic single particle excitations~\cite{Anderson-1958}. Therefore, a dirty, noninteracting CM in three dimensions only supports two phases, namely a diffusive metal and a localized Anderson insulator (AI)~\cite{Abrahams-1979}. The Anderson localization is a continuous quantum phase transition (QPT), for which the spatial dimension $d=2$ serves as the lower critical dimension~\cite{Abrahams-1979}. This QPT is reflected in the spatial variation of the wave function and can only be 
captured by 
quantities 
that 
probe its extended or localized nature, such as the inverse participation ratio (IPR)~\cite{Bell-1970,Wegner-1980}, and the typical DOS (TDOS)
~\cite{Janssen-1998,Dobrosavljevic-2003,Schubert-2010}. However, a self averaging quantity such as the average DOS (ADOS) itself remains unaffected by the localization transition.

In contrast to CMs, a half-filled (i.e. undoped) three dimensional DSM is an incompressible fermionic quantum critical system,
which leads to a quadratically vanishing DOS at the Fermi level. Consequently, any weak disorder
 acts as an irrelevant perturbation for a three dimensional DSM. Therefore, the three dimensional ballistic DSM phase should remain stable up to a critical strength of disorder $W_c$. At this threshold, the DSM undergoes a continuous QPT into a compressible diffusive metal (CDM) phase~\cite{Fradkin-1986,Goswami-2011}, and the ADOS at zero energy acts as an order parameter for describing this transition, while distinguishing it from a conventional Anderson localization. Based on a nonlinear sigma model analysis, the CDM phase has been argued to display an Anderson localization at a higher strength of disorder $W_l$, which belongs to the unitary (A) universality class~\cite{Fradkin-1986}. The first analysis of the DSM-CDM transition in Ref.~\onlinecite{Fradkin-1986} was a mean-field one, and
 only recently
 the non-Gaussian nature of this quantum critical point (QCP) has been elucidated~\cite{Goswami-2011}.
  Subsequently, the DSM-CDM phase transition has been addressed 
  using various analytic~\cite{Ominato-2014,Bitan-2014, Leo-2015, Sergey-2015} and numerical methods~\cite{Kobayashi-2014,Brouwer-2014,Pixley-2015}.
  However, the effects of large disorder and the Anderson localization transition have not been studied through any numerically exact approach.  Moreover, there have been recent claims that rare-region effects convert the DSM into a CDM for an infinitesimally weak disorder strength~\cite{Nandkishore-2014} so that there is in fact no DSM-CDM transition at all!
  Therefore, it is still an important open question to figure out how many distinct quantum phases indeed exist in a dirty 3D Dirac system for finite disorder (one or two or three with DSM/CDM/AI being all the possibilities) through a non-perturbative calculation, in addition to establishing the appropriate universality of any applicable disorder-driven QPT in the system.  In the current paper we answer this question through extensive exact numerical work supplemented by theoretical arguments. 

\begin{figure}[htb]
\centering
\begin{minipage}{.25\textwidth}
  \centering
  \includegraphics[width=0.7\linewidth,angle=-90]{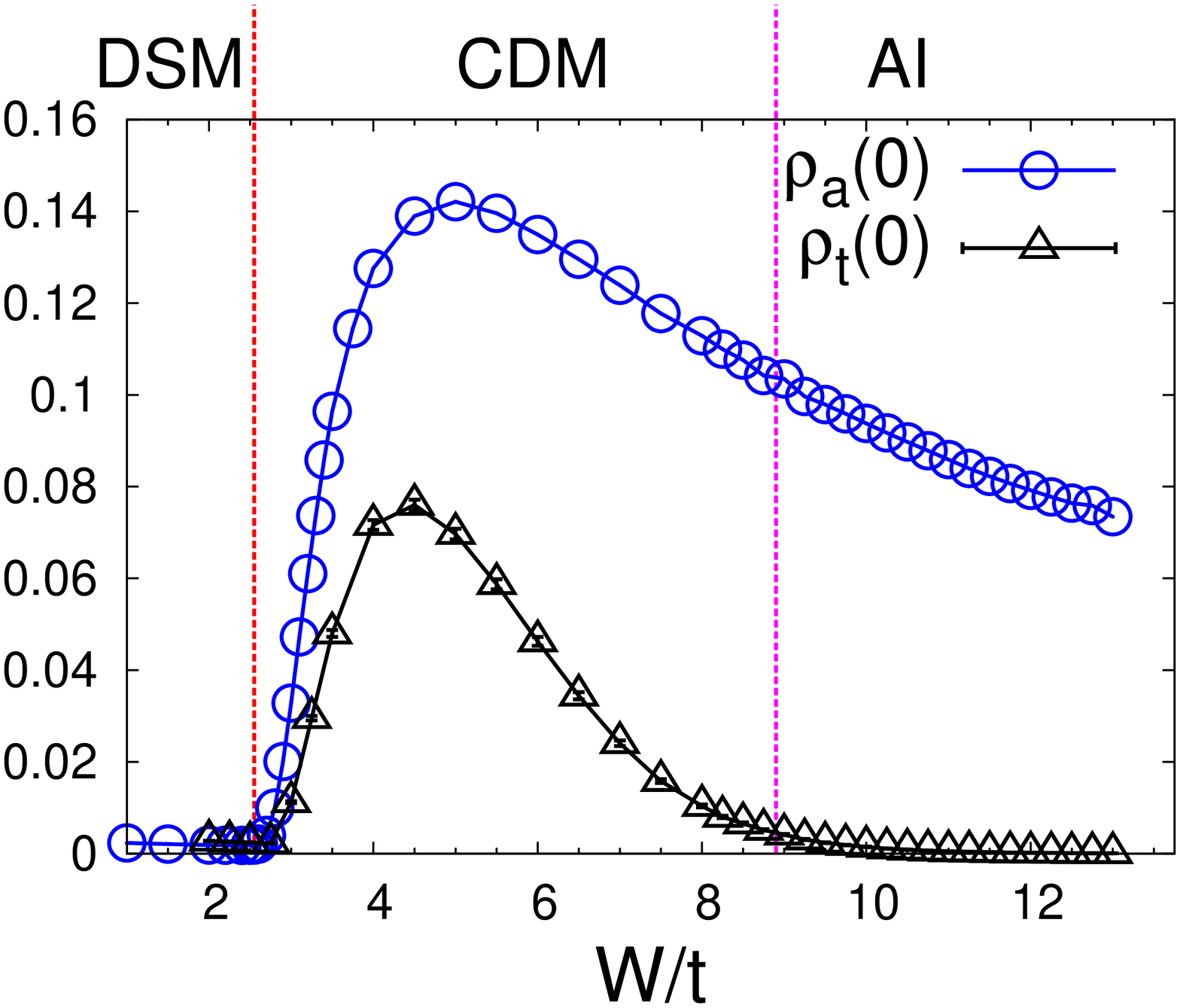}
\end{minipage}%
\begin{minipage}{.25\textwidth}
  \centering
  \includegraphics[width=0.9\linewidth]{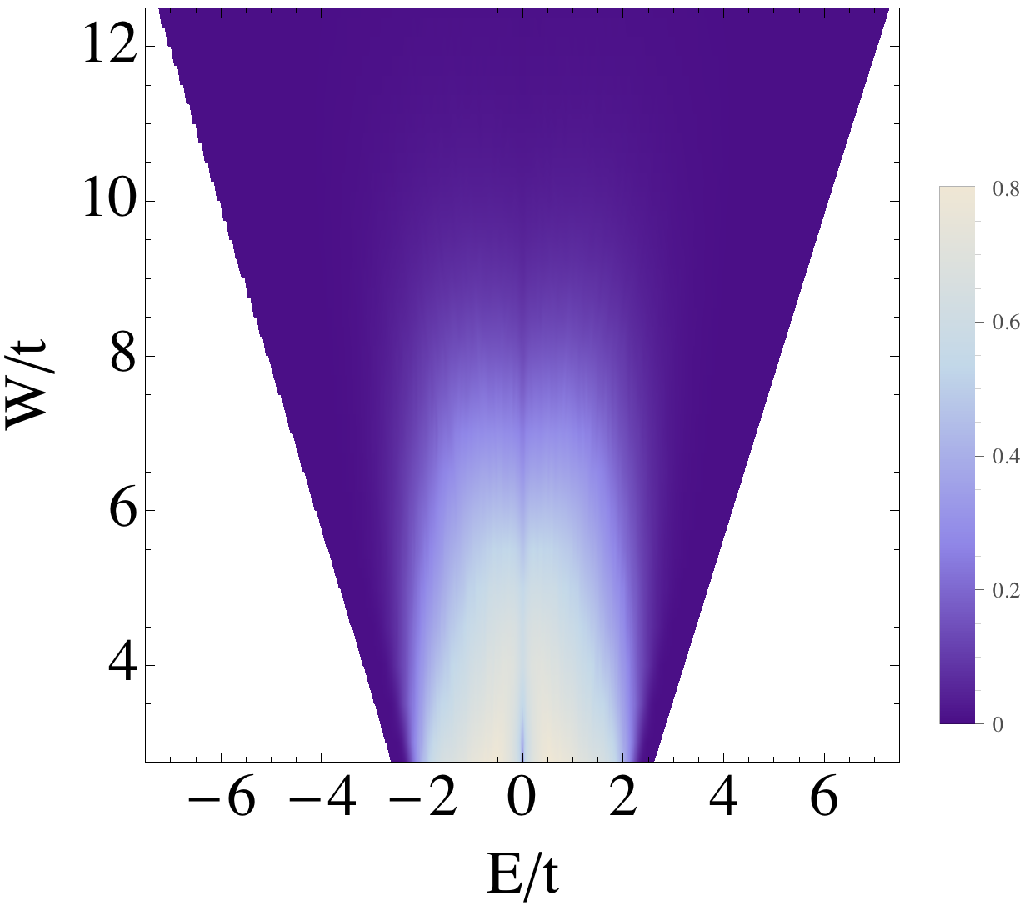}
\end{minipage}
\caption{Phase diagram of three dimensional Dirac semimetals. (Left) $\rho_a(0)$ and $\rho_t(0)$ for $L=60$, clearly displaying the existence of three distinct phases DSM, CDM, and AI. The TDOS 
tracks the ADOS 
inside the DSM, through the semimetal-metal QCP, 
and then $\rho_t(0)$ goes to zero at $W_l$, signaling the Anderson localization transition.
(Right)
$M_{edge}(E)$ as a function of energy and disorder strength for $L=60$. The white and the blue regions are respectively metallic and localized, and the blank region is gapped (outside the bandwidth) which clearly show the existence of the energy dependent mobility edge. The value of $M_{edge}(E)$ is shown in the key. }
 \label{fig:rhozero}
\end{figure}

We study the phase diagram of a 
dirty DSM in three dimensions using numerically exact methods. We calculate the ADOS and the TDOS on sufficiently large lattice sizes reaching up to $60^3$ sites, using the
kernel polynomial method (KPM)~\cite{Weisse-2006}. As the TDOS tracks the ADOS in any metallic phase and also serves as an order parameter for Anderson localization~\cite{Janssen-1998,Dobrosavljevic-2003}, it can naturally capture both possible (DSM-CDM-AI) QPTs (see Fig.~\ref{fig:rhozero}). 
We firmly establish that a disordered DSM in three dimensions possesses two distinct QPTs as a function of disorder strength, as shown in Figs.~\ref{fig:rhozero} and~\ref{fig:dostE}. The TDOS at zero energy is only finite in the CDM region $W_c<W<W_l$, in contrast to the ADOS at zero energy which is finite for any $W>W_c$. This provides unambiguous evidence for the the existence of three phases and the different nature of the two distinct disorder-tuned QPTs. The nature of the QCP between the DSM and the CDM phases has been studied through an extensive numerical computation of the ADOS in Refs.~\cite{Kobayashi-2014,Pixley-2015}. As the TDOS and the ADOS track each other in a metal, our calculation provides further numerical evidence for the stability of the DSM phase in the presence of weak disorder.
We supplement the DOS analysis through exact calculations of the wave function for small system sizes in order to estimate the critical exponents of the Anderson localization transition. Here, our goal is not to establish the precise values for the critical exponents of the Anderson localization transition. Rather, we want to show that the estimated critical exponents for the CDM-AI QCP are comparable to the ones known for the conventional orthogonal (AI) Wigner-Dyson universality class~\cite{Belitz-1994,Evers-2008}, and are strikingly different from the ones obtained for the DSM-CDM QCP~\cite{Kobayashi-2014,Pixley-2015}. Since our model of a DSM preserves time reversal symmetry, we expect the Anderson localization transition to be described by the orthogonal (AI) class~\cite{Belitz-1994,Evers-2008,supp} for short range disorder. 

We consider the following Hamiltonian on a cubic lattice with periodic boundary conditions,
\begin{equation}
H=\frac{1}{2}\sum_{{\bf r},\hat{\mu}}\left(i t \psi_{{\bf r}}^{\dag}\,\alpha_{\mu}\psi_{{\bf r}+{\bf e}_{\hat{\mu}}}  + \mathrm{H.c}\right)+\sum_{{\bf r}} V({\bf r})\psi_{{\bf r}}^{\dag}A_W\psi_{{\bf r}}
\label{eqn:ham}
\end{equation}
where $\psi_{{\bf r}}^T=(c_{{\bf r},+,\uparrow},c_{{\bf r},-,\uparrow},c_{{\bf r},+,\downarrow},c_{{\bf r},-,\downarrow})$ is a four component Dirac spinor composed of an electron at site ${\bf r}$, with a parity ($\pm$) and spin ($\uparrow/\downarrow$),
$\hat{\mu}=\hat{x},\hat{y},\hat{z}$, ${\bf e}_{\hat{\mu}}$ is a unit vector that points to each nearest neighbor, and the $\alpha_{\mu}$ are $4 \times 4$ anticommuting Dirac matrices. In the absence of disorder, the tight binding model gives rise to eight Dirac cones at the $\Gamma$, $M$, $R$ and $X$ points of the cubic Brillouin zone. The disorder potential at each site $V({{\bf r}})$ is a randomly distributed variable between $[-W/2,W/2]$ and $A_W$ is a matrix that specifies the type of disorder. We primarily focus on the random axial chemical potential $A_W = \gamma_5=i\alpha_1\alpha_2\alpha_3$, but also discuss the effects of a random scalar potential $A_W=I_{4\times4}$ (where $I_{4\times4}$ is a four by four identity matrix). We consider these two types of disorder because we want to demonstrate universality in the sense that both disorders give rise to identical phase diagrams.
 This is also implied by the existence of two continuous U(1) symmetries under $\psi \to e^{i \theta} \psi$ and $\psi \to e^{i\phi \gamma_5} \psi$. In addition, axial disorder is pertinent for addressing the phase diagram of dirty gapless superconductors.


\emph{TDOS and ADOS:} 
To track the nature of each phase of the model defined in Eq. (\ref{eqn:ham}), we first calculate the ADOS, the local DOS (LDOS), and the TDOS using the KPM (see Ref.~\onlinecite{Weisse-2006} for details), which are defined as
\begin{eqnarray}
\rho_a(E) &=& \left\langle \frac{1}{4V}\sum_{i=1}^{V}\sum_{\alpha=1}^4\delta(E-E_{i\alpha}) \right\rangle,
\\
\rho_{i\alpha}(E) &=& \sum_{k,\beta}|\langle k,\beta |i, \alpha \rangle |^2\delta(E-E_{k\beta}),
\\
\rho_{t}(E) &=& \exp\left(\frac{1}{4N_s }\sum_{i=1}^{N_s}\sum_{\alpha=1}^4 \Big\langle \log \rho_{i\alpha}(E) \Big\rangle\right).
\end{eqnarray}
The size of the system considered is $V=L^3$, where $ |i, \alpha \rangle $ denotes an eigenstate at site $i$ and orbital $\alpha$ (one of the four Dirac orbitals) with an energy $E_{i\alpha}$, and the $\langle \dots \rangle$ denotes a disorder average. As the ADOS is self averaging, we only include the disorder average to smoothen the data. For the calculations of the ADOS presented here with a system size $V=60^3$ we have only used $12$ disorder averages. In contrast, the TDOS is not a self averaging quantity and it requires a significant number of disorder averages to obtain a convergent result. After the disorder averaging, translation symmetry is  restored and consequently we also sum over a finite number of lattice sites $N_s\ll V$ to improve the statistics. We provide the parameters used in the study in the Supplemental Material~\cite{supp}.
Central to the KPM, we first expand the LDOS or the ADOS in terms of the orthogonal Chebyshev polynomials up to an order $N_c$. For the ADOS we have considered $N_c=1028$, whereas the computation of the TDOS is much more demanding~\cite{Weisse-2006} and for most of the calculations we have used $N_c = 8192$, unless stated otherwise.

\begin{figure}[htb]
\centering
\begin{minipage}{.25\textwidth}
  \centering
  \includegraphics[width=0.7\linewidth,angle=-90]{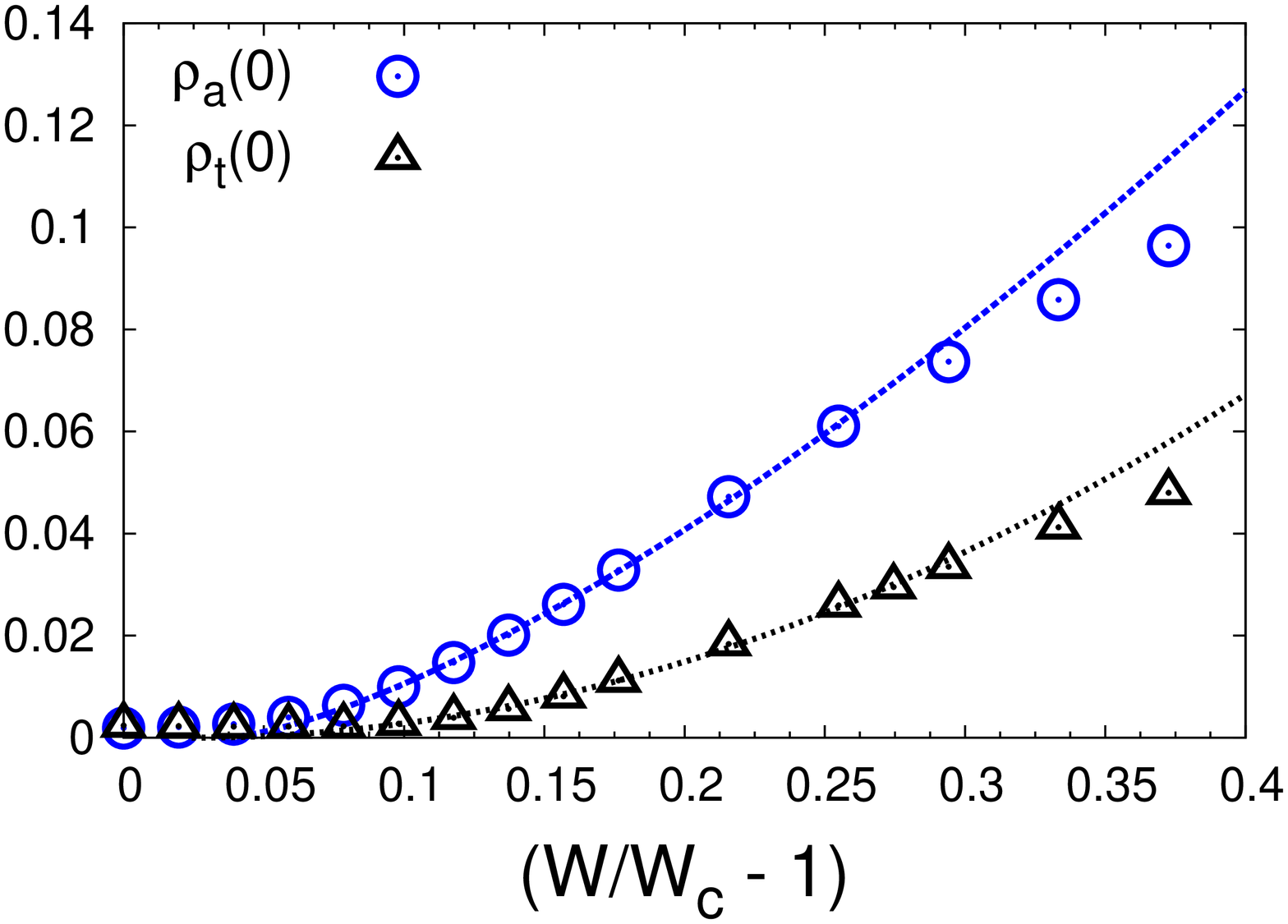}
\end{minipage}%
\begin{minipage}{.25\textwidth}
  \centering
  \includegraphics[width=0.7\linewidth,angle=-90]{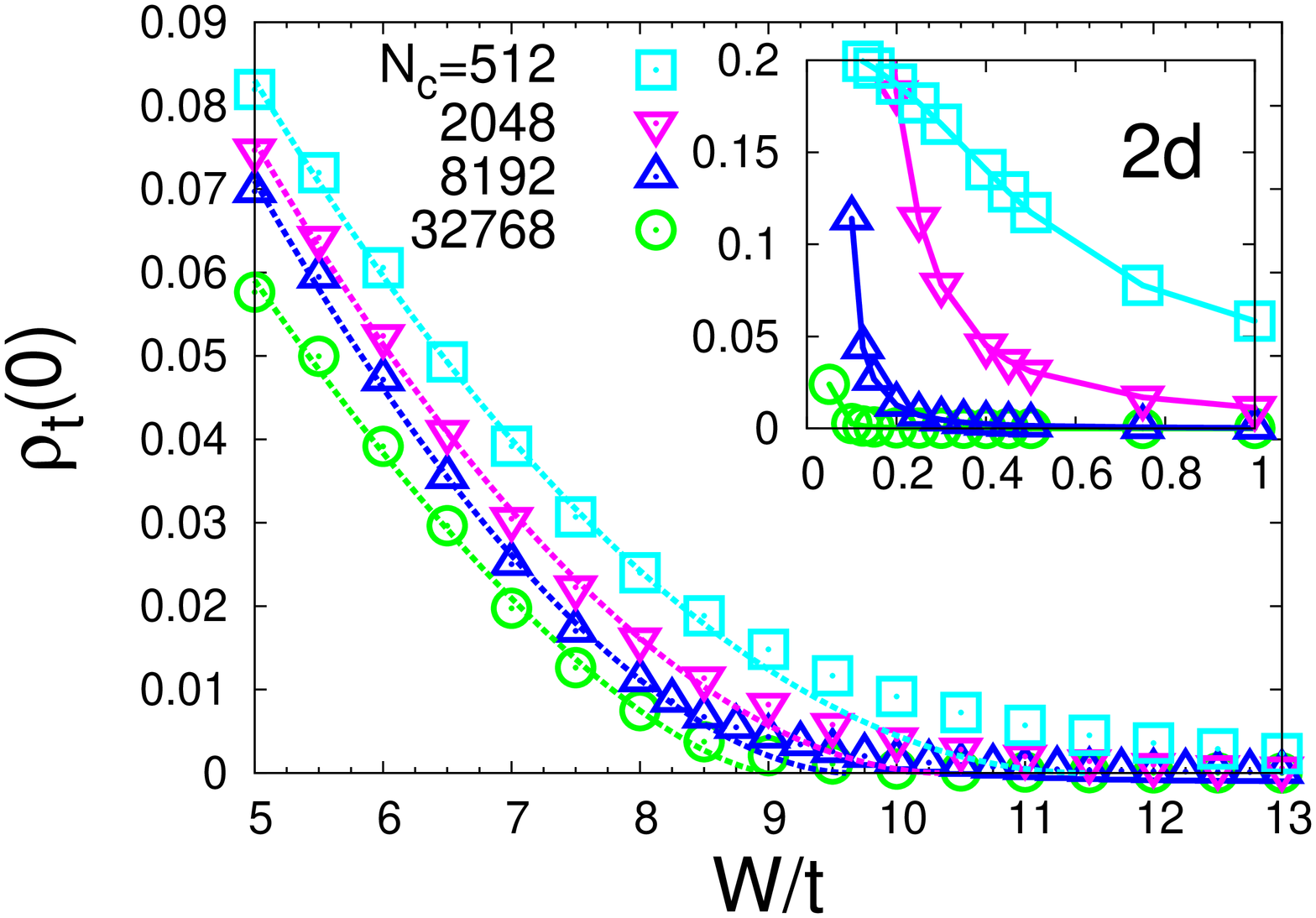}
\end{minipage}
\caption{Critical properties of the two quantum critical points. (Left) Power law dependence of $\rho_a(0)$ and $\rho_t(0)$ as a function of disorder strength in the vicinity of the DSM-CDM QCP for $L=60$. (Right)
The Anderson localization QCP from the TDOS as a function of the Chebyshev expansion order $N_c$ for $L=30$ respectively in three  and (Inset) two dimensions. We find the localization transition in three dimensions is converging to $W^{3d}_l/t = 8.9\pm 0.3$ in three dimensions whereas  in two dimensions $W^{2d}_l/t\rightarrow 0$ as $N_c\rightarrow\infty$
by extrapolating $\rho_t(0)$ to zero (dashed lines). }
\label{fig:rho2d}
\end{figure}

The DSM is characterized by an ADOS $\rho_a(E) \propto |E|^2$, that vanishes at zero energy. As shown in Fig.~\ref{fig:rhozero}, both $\rho_a(0)$ and $\rho_t(0)$ vanish for weak disorder strengths, and concomitantly become finite after passing through the DSM-CDM QCP at $W_c/t \approx 2.55$~\cite{Pixley-2015}, in a power law fashion as
\begin{eqnarray}
\rho_a(E=0,W \ge W_c) &\sim& (W-W_c)^x,
\\
\rho_t(E=0,W \ge W_c) &\sim& (W-W_c)^{x_t},
\end{eqnarray}
where we find $x=1.4 \pm 0.2$ (in agreement with Refs.~\onlinecite{Kobayashi-2014,Pixley-2015} for much larger system sizes) and $x_t = 2.0 \pm 0.3$ as shown in Fig.~\ref{fig:rhozero}.  Physically, the average moments of the LDOS $\langle \rho_i(E)^n \rangle$ captures the multifractal nature of the disordered wavefunction~\cite{Belitz-1994,Janssen-1998,Foster-2012}.
Therefore the difference $x_t-x=0.60\pm 0.36$ reveals the underlying multifractal nature of the DSM-CDM QCP. We have checked that the value of $W_c$ determined from $\rho_a(0)$ is within numerical accuracy unaffected by increasing $N_c$.

Inside the CDM phase $\rho_t(0)$ tracks $\rho_a(0)$ ($\rho_t \sim \rho_a$) up to $W/t \approx 4.5$. As $\rho_a$ is a \emph{self averaging} quantity, we can conclude that the DSM-CDM transition 
completely independent of
Anderson localization.
For larger values of disorder ($W/t > 4.5$) $\rho_a$ remains finite and $\rho_t$ goes to zero continuously at the localization transition as
\begin{equation}
\rho_t(E=0,W\le W_l) \sim (W_l-W)^{\beta}.
\end{equation}
We determine the location of the transition by studying the effect of increasing the Chebyshev expansion order on $\rho_t$, as $W_l$ has been shown to be sensitive to $N_c$ in Ref.~\onlinecite{Schubert-2005}. By extrapolating $\rho_t(E=0)$ to zero, we find $W_l/t = 8.9\pm 0.3$ with an order parameter critical exponent $\beta =1.5\pm0.2$, as shown in Fig.~\ref{fig:rho2d}.


\begin{figure}[htb]
\centering
\begin{minipage}{.25\textwidth}
  \centering
  \includegraphics[width=0.9\linewidth]{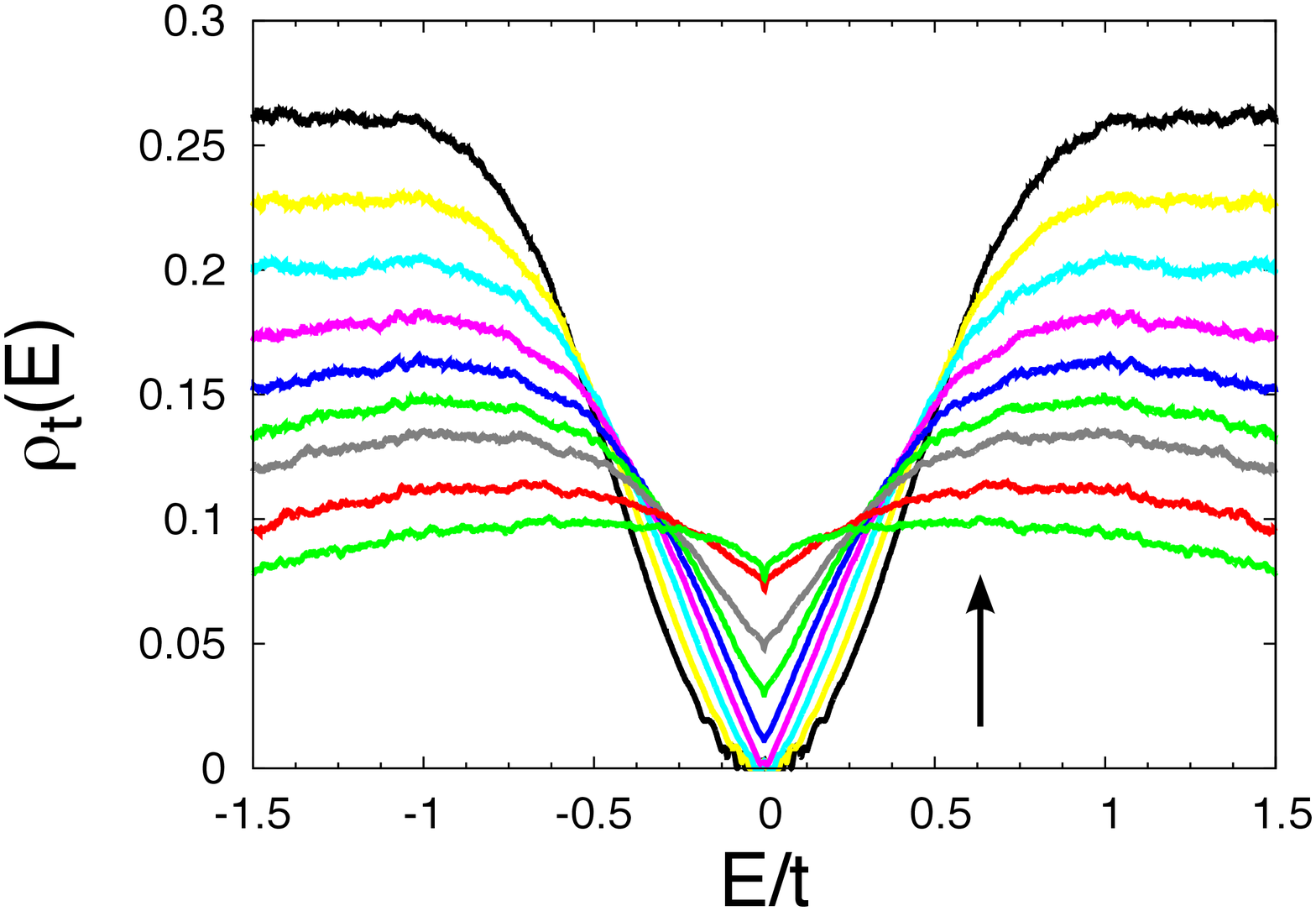}
\end{minipage}%
\begin{minipage}{.25\textwidth}
  \centering
  \includegraphics[width=0.9\linewidth]{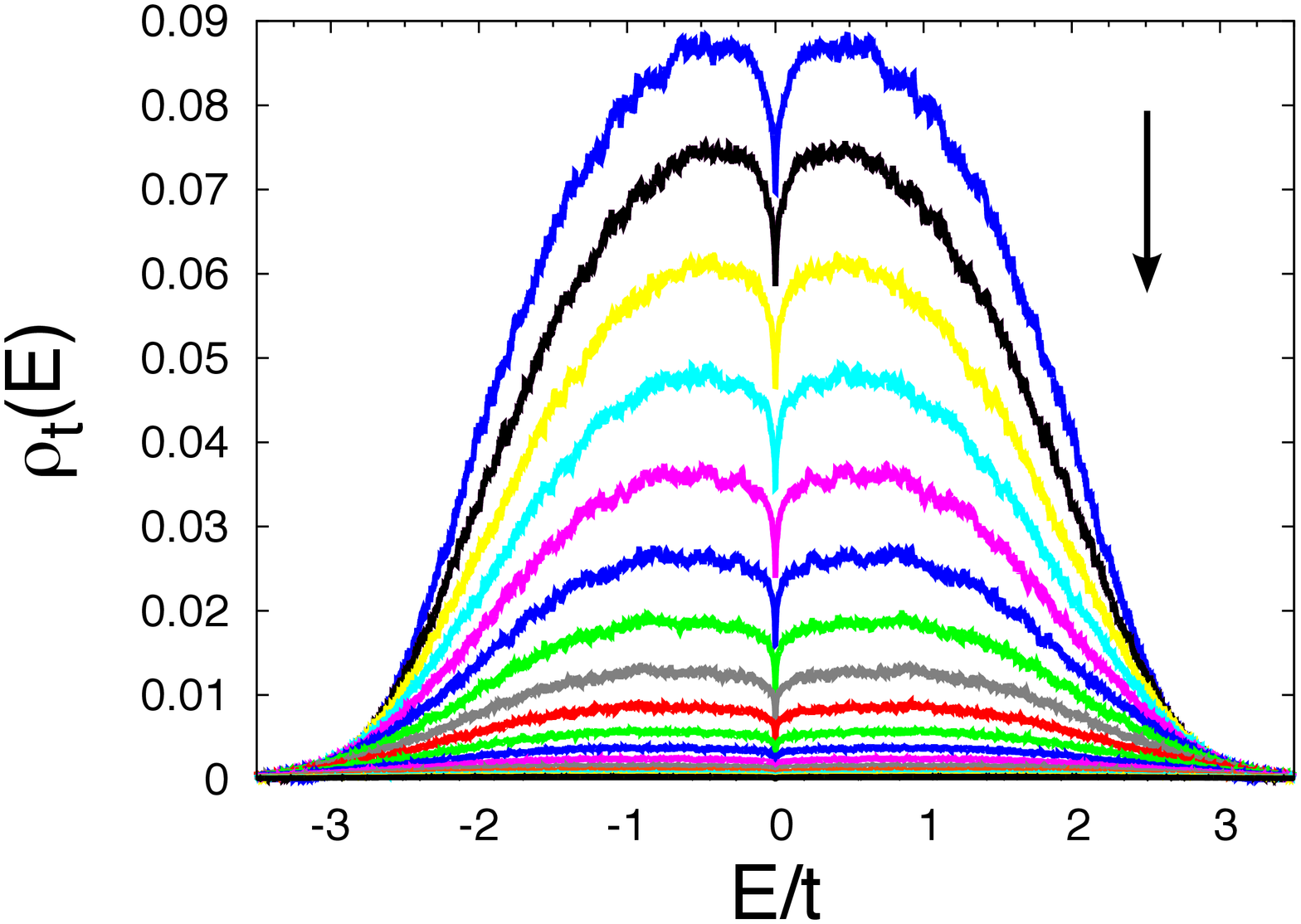}
\end{minipage}
\caption{TDOS with $L=60$ as a function of energy for various values of disorder. (Left) Passing through the DSM-CDM QCP, (Right) passing through the Anderson localization transition. The arrows denote increasing disorder strengths from $W/t=2.0$ to $4.5$ (Left) and from $W/t=5.0$ to $13.0$ (Right). We see no signs of rare region effects~\cite{Nandkishore-2014} in the numerics.}
 \label{fig:dostE}
\end{figure}

In order to further understand the stability of the CDM against localization effects, we compare the TDOS for a DSM in two and three spatial dimensions in Fig.~\ref{fig:rho2d}. The two dimensional model of a DSM is obtained by setting the hopping along the $z$ direction to zero in Eq. (\ref{eqn:ham}), which gives rise to four Dirac cones at the high symmetry points of the Brillouin zone for a square lattice. In two dimensions, both axial and scalar potential disorders are marginally relevant perturbations, and an infinitesimal amount of disorder leads to an AI~\cite{Aleiner-2006,Altland-2006,supp}. This is in excellent agreement with our numerical results, namely $W_l^{2d}/t\rightarrow 0$ as $N_c\rightarrow \infty$.  In contrast, in three dimensions we find the DSM phase to be stable up to $W_c$ (Fig.~\ref{fig:rhozero}), where $\rho_a(0)$ and $\rho_t(0)$ simultaneously develop finite values, and the CDM phase undergoes an Anderson localization at a much larger disorder strength $W_l$. Therefore, our results 
establish the existence of three distinct ground states for a dirty DSM in three dimensions (DSM/CDM/AI with increasing disorder), 
 in
agreement with the
field theoretic analysis.

By considering the energy dependence of $\rho_t$ as shown in Fig.~\ref{fig:dostE}, we find $\rho_t(E) \sim |E|^2$ inside the DSM phase (i.e. in agreement with $\rho_a(E)$~\cite{supp}). In contrast, $\rho_t(0)$ approaches a constant inside the CDM phase. Within this framework, we can
capture the location of the mobility edge separating the CDM and AI phases by considering the ratio $M_{\mathrm{edge}}(E)=\rho_t(E)/\rho_a(E)$. As shown in Fig.~\ref{fig:rhozero}, starting from the CDM phase,
the system develops a finite, energy dependent mobility edge and the entire band localizes for a larger disorder strength going to an AI phase, qualitatively similar to the situation for a three dimensional CM. 
Due to the underlying Dirac band structure, the shape of the mobility edge is different from a conventional tight binding model (e.g. see Ref.~\cite{Brndiar-2006}). This phase diagram also demonstrates that a DSM away from half-filling behaves as a CM.

\emph{Wave function:}
In order to understand the critical properties of the localization transition and compare our results with the well established universality classes for Anderson localization, we now study the properties of the wave function.  The qualitative nature of the wave function also displays all of the physics we have discussed so far~\cite{supp}. 
Here, we focus on
the average participation ratio defined as
\begin{equation}
P_{avg} = \left\langle \frac{[\sum_{i,\alpha} |\psi_{\alpha}({\bf r}_i)|^2]^2}{\sum_{i,\alpha} |\psi_{\alpha}({\bf r}_i)|^4 }\right\rangle
\end{equation}
for wave functions $\psi_{\alpha}({\bf r}_i)$ and only focus on the center of the band $E=0$ (otherwise $P^{-1}_{avg}$ will be a function of energy).  It is well known~\cite{Janssen-1998} that the IPR scales as $P^{-1}_{avg} \sim 1/V$ in a metallic phase and $P^{-1}_{avg} \sim \mathrm{const.}$  in a localized phase~\cite{supp}.
As these calculations require an exact diagonalization of the Hamiltonian, we are restricted to much smaller lattice sizes than considered for the DOS.  Here we focus on linear system sizes $L=4,6,8,10,12$ with $10,000$ disorder realizations for $L=4-10$ and $1,000$ for $L=12$. Due to the limited system sizes we first expand the IPR in the vicinity of the localization transition in terms of an unknown scaling function $f$ and $L$ dependent corrections,~\cite{Janssen-1998,Slevin-1999,Mirlin-2000,Mildenberger-2002,Brndiar-2006}
$
P^{-1}_{avg} \approx L^{-d_2}\left(f((1-W_l/W)L^{1/\nu}) + A_0/L^y\right).
$
We have introduced the fractal dimension $d_2$, the localization length exponent $\nu$, $A_0$ is a finite size correction, and $y$ the leading order irrelevant variable's critical exponent~\cite{Slevin-1999}.
To determine the location of the transition we calculate the the best recursive fit
to
 $Y_{IPR} = P^{-1}_{avg}L^{d_2} - A_0 L^{-y}$, using finite size scaling techniques~\cite{Slevin-1999,Murphy-2011}, see Fig.~\ref{fig:IPR}.  As a result, we find a localization transition at $W_l/t = 8.8 \pm 0.3$ in excellent agreement with the typical DOS results.  This also yields $d_2 = 1.2 \pm0.3$, $y =2.3\pm0.5$, and $A_0=1.8\pm0.6$.

Now that we have determined the critical point using two different methods we perform scaling data collapse to extract the localization length exponent,
which yields $\nu=1.5 \pm 0.2$. Within the numerical accuracy, $\nu$, $y$, and $d_2$ are consistent with the known exponents for the orthogonal (AI) class ($\nu =1.57$, $y=2.8$, and $d_2=1.23$)~\cite{Slevin-1999,Brndiar-2006,Varga-2015}. 
In previous studies of Anderson localization, it was found that typical DOS exponent satisfies $\beta=\nu (\alpha_0-d)$~\cite{Janssen-1998,Evers-2008}. We also find this relation to hold with $\alpha_0=4.0 \pm 0.3$, which is consistent with $\alpha_0=4.043$~\cite{Varga-2015} for class AI .   

\begin{figure}
\centering
\begin{minipage}{.25\textwidth}
  \centering
  \includegraphics[width=0.9\linewidth]{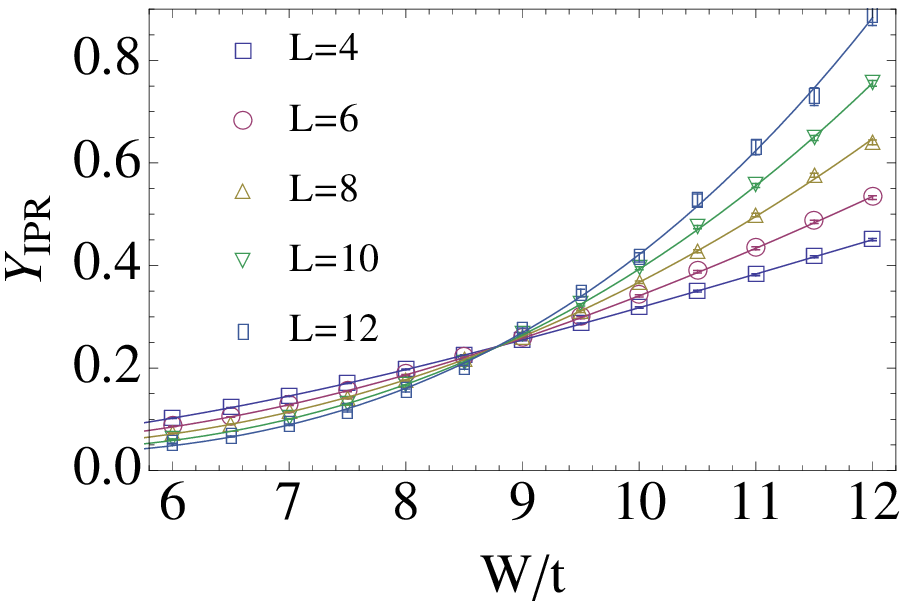}
\end{minipage}%
\begin{minipage}{.25\textwidth}
  \centering
  \includegraphics[width=0.9\linewidth]{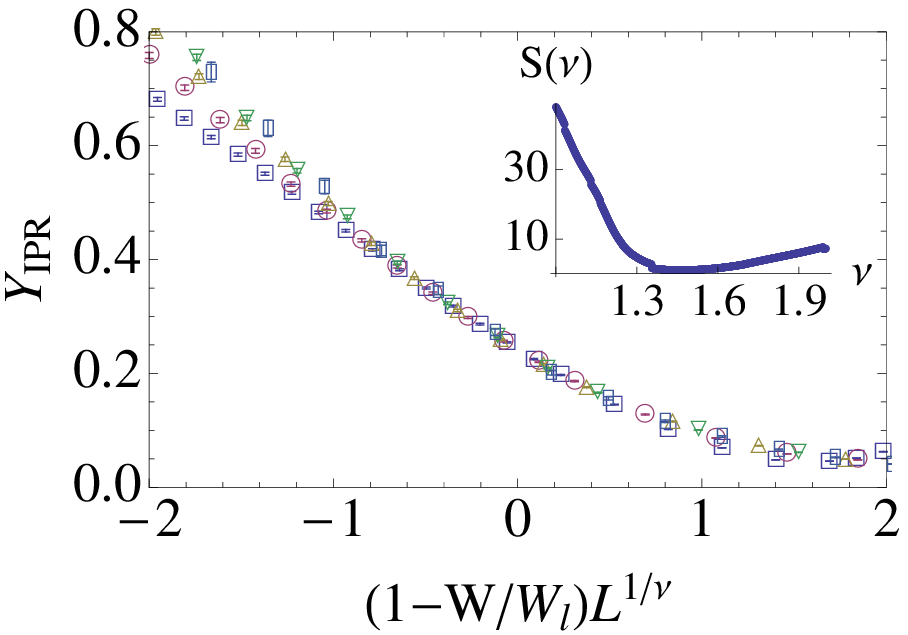}
\end{minipage}
\caption{Finite size scaling of the IPR.  (Left) Determining the location of the Anderson localization transition by optimizing the crossing location of $Y_{IPR}$ by third order polynomial fits to the IPR, this yields $W_l/t = 8.8\pm0.3$.  (Right) Data collapse of $Y_{IPR}$ in the vicinity of $W_l$ as a function of $(1-W_l/W)L^{1/\nu}$, yields a localization length exponent $\nu= 1.5\pm0.2$. (Inset) Optimizing the data collapse using the local linearity function $S(\nu)$ (as described in Ref.~\cite{Kawashima-1993}), the value of $\nu$ is determined by the minimum of $S(\nu)$. }
 \label{fig:IPR}
\end{figure}

In conclusion, we have performed a detailed numerically exact study of the quantum phase diagram and Anderson localization in three dimensional dirty DSMs. By studying the TDOS on sufficiently large lattices we have established the existence of three phases for a half-filled system DSM, CDM and AI,
separated by two different QCPs as a function of disorder.
We have supplemented this with a precise finite size scaling analysis of the IPR and obtained the critical exponents of the localization transition.
We also establish clear signatures for multifractility at the QPTs.
In addition, we find the localization properties for both axial and scalar potential disorders to be identical (to within numerically accuracy)~\cite{supp}, 
and therefore, we conclude that the QPTs driven by either disorder belong to the same universality class. As the model we have studied for axial or potential disorder separates into two blocks of Weyl Hamiltonians~\cite{supp}, our results are equally valid for three dimensional time reversal symmetric Weyl systems.

\emph{Acknowledgements:} We thank S. A. Kivelson, M. Foster, and S. Sachdev for useful discussions. This work is supported by JQI-NSF-PFC and LPS-CMTC. 
We acknowledge the University of Maryland supercomputing resources (http://www.it.umd.edu/hpcc) made available in conducting the research reported in this paper.

\bibliography{DSM_loc}

\pagebreak
\newpage
\onecolumngrid

\setcounter{figure}{0}
\makeatletter
\renewcommand{\thefigure}{S\@arabic\c@figure}
\setcounter{equation}{0} \makeatletter
\renewcommand \theequation{S\@arabic\c@equation}
\renewcommand \thetable{S\@arabic\c@table}

\maketitle

\section*{\Large{Supplemental Material}}
In the supplemental material we first provide field theoretic arguments to justify our claims regarding the universality class of the localization transition.
We then describe in detail the parameters used in the numerical study, present the typical DOS for axial and potential disorder, the system size dependence of $\rho_t(0)$, the finite size scaling of the IPR for potential disorder, present a comparison between the typical and average density of states in each phase studied in this work, and show both the inverse participation ratio for each system size studied and the wave function in each phase as a function of the distance to its maximal absolute value.  
 
\section{\textbf{Field theory arguments}}

In momentum space, the tight-binding Dirac Hamiltonian (Eq. (1) in the main text) in the absence of disorder becomes
\begin{equation}
H=t \sum_{\mu=1}^{3}\sin k_\mu \alpha_\mu
\end{equation}
with the dispersion relation
\begin{equation}
E_{\mathbf{k}}=\pm t \; \sqrt{\sin^2k_x+\sin^2 k_y+\sin^2 k_z}. 
\end{equation} The conduction and the valence bands touch at following eight high symmetry points (i) $(0,0,0)$, (ii) $(\pi, \pi,0)$, (iii) $(\pi,0,\pi)$, (iv) $(0,\pi, \pi)$, (v) $(\pi,0,0)$, (vi) $(0,\pi,0)$, (vii) $(0,0,\pi)$, and (viii) $(\pi, \pi,\pi)$ of the cubic Brillouin zone, as a consequence of the Fermion doubling theorem. The lattice Hamiltonian commutes with $\gamma_5=i\alpha_1 \alpha_2 \alpha_3$, which generates a $U(1)$ chiral symmetry. Due to the absence of any mass term, the Hamiltonian naturally anti-commutes with two other Dirac matrices $\beta$ and $\beta \gamma_5$. The linearized Dirac Hamiltonian around these points can be written as
\begin{equation}
H_{D,j}=v_{1,j} k_x \alpha_1 +v_{2,j} k_y \alpha_2 +v_{3,j} k_z \alpha_3. 
\end{equation} 
The velocities $(v_1,v_2,v_3)$ at the eight Dirac points are respectively given by $(v,v,v)$, $(-v,-v,v)$, $(-v,v,-v)$, $(v,-v,-v)$, $(-v,v,v)$, $(v,-v,v)$, $(v,v,-v)$ and $(-v,-v,-v)$, where $v=t$ (in units of $\hbar=a=1$ for a lattice spacing $a$). With suitable unitary transformations at different cones we can covert all of the linearized Hamiltonians at cones (ii) to (viii) into the same form as that of (i). For example the spinor $\psi_8$ at $(\pi, \pi, \pi)$ can be transformed as $\beta \psi_8$ or $i\beta \gamma_5 \psi_8$, to arrive at the desired form. Thus, the linearized theory will have an emergent $SU(8)$ flavor symmetry. This is a continuum approximation to the cubic point group symmetry. Similarly, if we set the hopping in the $z$ direction to zero, in order to obtain the two dimensional Hamiltonian studied in the main text, we obtain four species of four-component Dirac fermions in the vicinity of $(0,0)$, $(\pi, 0)$, $(0,\pi)$ and $(\pi, \pi)$. The linearized theory then demonstrates an $SU(4)$ flavor symmetry. Any inter-valley scattering process due to disorder and/or interaction usually reduces this emergent symmetry. 

The discussion of disorder effects can be easily understood in the chiral basis, with $\alpha_1=\sigma_1 \otimes \tau_3$, $\alpha_2=\sigma_2 \otimes \tau_3$, $\alpha_3=\sigma_3 \otimes \tau_3$, $\beta = \sigma_0 \otimes \tau_1$ and $\gamma_5=\sigma_0 \otimes \tau_3$. Here, $\sigma_\mu$ are Pauli matrices acting on the spin index, and $\tau_\mu$ are Pauli matrices which act on the chiral indices. In this basis the lattice Hamiltonian in the presence of both axial and potential disorders is block diagonal and consists of the following two $2 \times 2$ Weyl Hamiltonians 
\begin{eqnarray}
H_{1,W}=\frac{1}{2}\sum_{{\bf r},\hat{\mu}}\left(i t \chi_{1,{\bf r}}^{\dag}\,\sigma_{\mu}\chi_{1,{\bf r}+{\bf e}_{\hat{\mu}}} + \mathrm{H.c}\right)+\sum_{{\bf r}} [V_{ax}({\bf r})+V_{p}({\bf r})]\chi_{1,{\bf r}}^{\dag}\chi_{1,{\bf r}},  \\
H_{2,W}=-\frac{1}{2}\sum_{{\bf r},\hat{\mu}}\left(i t \chi_{2,{\bf r}}^{\dag}\,\sigma_{\mu}\chi_{2,{\bf r}+{\bf e}_{\hat{\mu}}} + \mathrm{H.c}\right)+\sum_{{\bf r}} [-V_{ax}({\bf r})+V_{p}({\bf r})]\chi_{2,{\bf r}}^{\dag}\chi_{2,{\bf r}}, 
\end{eqnarray} 
where $\chi^T_{1/2,{\bf r}}=(c_{{\bf r},1/2,\uparrow},c_{{\bf r},1/2,\downarrow})$ is a two component spinor composed of the linear combinations $c_{{\bf r},1,\sigma}=[c_{{\bf r},+,\sigma} + c_{{\bf r},-,\sigma}]/\sqrt{2}$ and $c_{{\bf r},2,\sigma}=[c_{{\bf r},+,\sigma} - c_{{\bf r},-,\sigma}]/\sqrt{2}$.
Each block by itself represents a model Hamiltonian of a Weyl semimetal on a non-centrosymmetric cubic lattice, which has four right handed and four left handed Weyl points~\cite{Goswami-2013}. For this reason axial and potential disorders show identical behavior, and the model clearly has time reversal symmetry. For each two component lattice model (in the presence of the intervalley scattering due to short range disorder), the relevant symmetry class is the orthogonal Wigner-Dyson class or AI. Consequently, the CDM to AI transition will belong to this class. Similar to the case of graphene with inter-valley scattering~\cite{Aleiner-2006,Altland-2006}, in the replica formalism the appropriate $Q$ matrix for this problem will be chosen from the space $\frac{Sp(4n)}{Sp(2n) \times Sp(2n)}$. In the presence of intervalley scattering (which is always a relevant term inside the diffusive metal phase) the $Q$ matrices for different Weyl cones are locked into a single one. In Ref.~\onlinecite{Fradkin-1986}, the $Q$ matrix was chosen to be a unitary matrix, which is actually applicable for a time reversal symmetry breaking Weyl semimetal. It is also interesting to recall that the identical behaviors of the axial and the scalar potential disorder for the DSM-CDM transition were first noted in the Supplementary Material of Ref.~\onlinecite{Goswami-2011}. The renormalization group flow with respect to the ballistic DSM fixed point in the presence of both types of disorder was found to be
\begin{eqnarray}
\frac{d\Delta_j}{dl}=-\Delta_j+2\Delta^2_j, \\
z=1+\sum_{j} \Delta_j,
\end{eqnarray} where $z$ is the scale dependent dynamic exponent, and $\Delta_A$ and $\Delta_{P}$ denote the dimensionless coupling constants for axial and potential disorder respectively. The DSM-CDM transition occurs along $\Delta_{c,A}+\Delta_{c,P}=1/2$, with $z_{c1}=3/2$ and $\nu_{c1}=1$. This further supports our symmetry based argument for the identical behaviors of the scalar potential and the axial disorders, which has also been verified through extensive numerical calculations~\cite{Pixley-2015}. Note that we are using the subscripts $1$ and $2$ to denote the DSM-CDM and CDM-AI quantum phase transitions respectively.

For completeness we also write down the following nonlinear sigma model for the 4-component Dirac Hamiltonian, 
\begin{eqnarray}
S[Q]=-\frac{\pi \rho_{a}(0) D(0)}{8} \int d^3 x \; \mathrm{Tr} (\nabla Q)^2 + \frac{\pi \rho_a(0)}{2} \int d^3 x \; \mathrm{Tr}[\Omega Q] ,
\end{eqnarray} where 
$\rho_a(0) \sim \delta^{\nu_{c1} (d-z_{c1})}$ is the average DOS inside the CDM phase, $D$ is the diffusion constant inside the CDM phase at the zero energy and $\delta=2(\Delta_A+\Delta_V-1/2)$ is the reduced distance from the DSM-CDM critical point. The product $\rho_a(0) D(0) \sim \sigma_{xx}(0) \sim \delta^{\nu_{c1}(d-2)}$, where $\sigma_{xx}$ is the longitudinal conductivity of the CDM. In the above equation $\Omega$ is the external frequency. The dynamic critical exponent for the Anderson localization is $z_{c2}=d=3$, which is clearly distinct from $z_{c1}=3/2$ for the DSM-CDM QCP. The beta function for this model is well known up to five loop order and can be found in Refs.~\onlinecite{Hikami-1981, Wegner-1989, Hikami-1992}.

\section{\textbf{Additional numerical results}}
\begin{table}[h]
\caption{Table of the parameters used for the typical DOS.  We denote the number of disorder realizations used as $N_D$ and the number of lattice sites that we average the typical DOS over is $N_{\mathrm{site}}$.}
\begin{center}
  \begin{tabular}{| c | c | c |}
    \hline
    L & $N_D$ & $ N_{\mathrm{site}} $   \\ \hline 
   20 & $  2000 $ & $20$ \\ \hline
    30 & $  800 $ & $10$ \\ \hline
     40 & $  700 $ & $10$ \\ \hline
      50 & $  600 $ & $10$ \\ \hline
       60 & $  520 $ & $10$ \\ 
    \hline
  \end{tabular}
\end{center}
\end{table}

\begin{figure}[htb]
\begin{minipage}{.3\textwidth}
  \centering
  \includegraphics[width=0.9\linewidth]{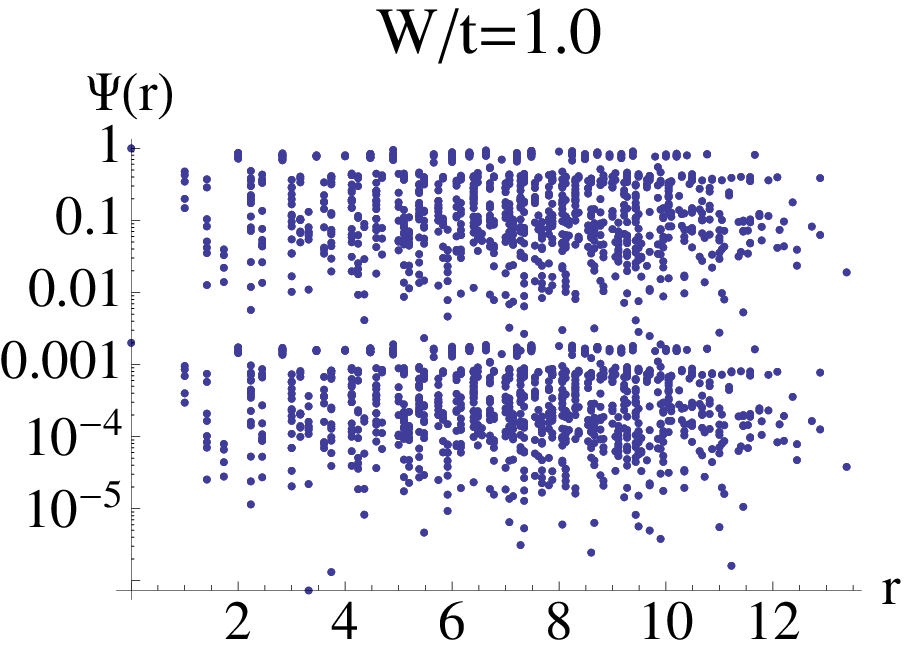}
\end{minipage}%
\begin{minipage}{.3\textwidth}
  \centering
  \includegraphics[width=0.9\linewidth]{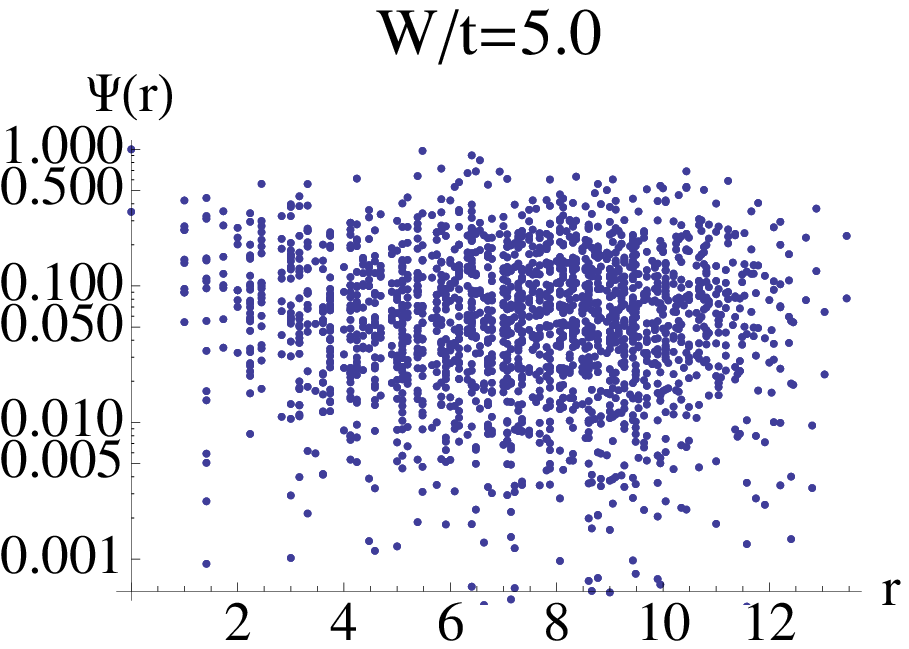}
\end{minipage}
\begin{minipage}{.3\textwidth}
  \centering
  \includegraphics[width=0.9\linewidth]{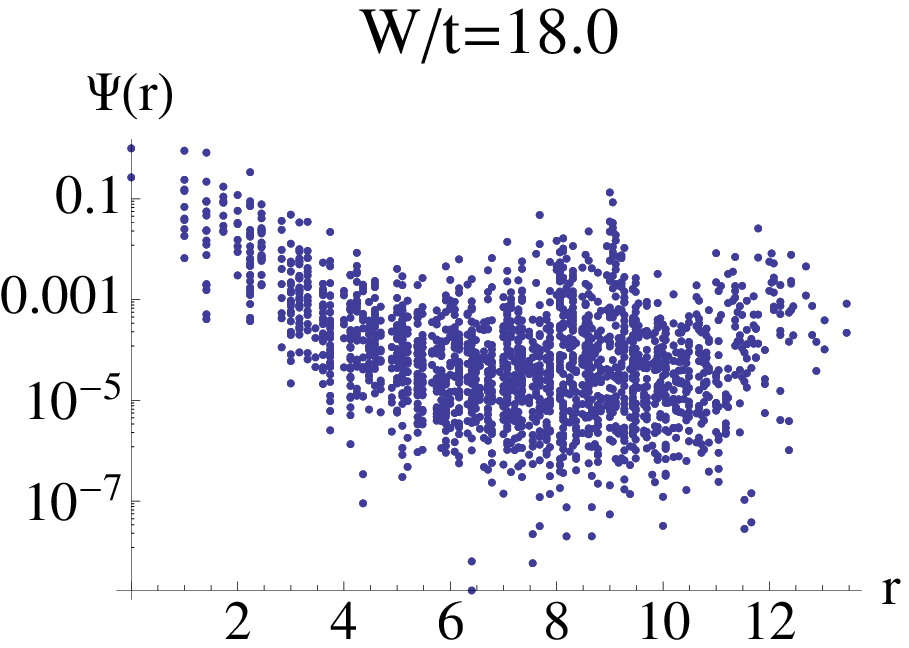}
\end{minipage}%
\caption{Decay of the magnitude of the wave function $\Psi(r) = |\psi(|{\bf r} - {\bf r}_{\mathrm{max}}|)|/|\psi({\bf r}_{\mathrm{max}})|$ for the center of the band as  a function of the distance to its maximal value $|\psi({\bf r}_{\mathrm{max}})|$ for $L=10$ in each phase studied.  We find no real qualitative change in the wave function across the DSM to CDM QCP (left and center), whereas in the Anderson insulating phase the wave function displays a clear decay (right).}
 \label{fig:WFall}
\end{figure}

The overall behavior of the wave function also captures the localization transition, while it is unaffected by the DSM-CDM QCP, see Fig.~\ref{fig:WFall}. We compute the decay of the wave function from its maximal value $\Psi({\bf r}) = |\psi(|{\bf r} - {\bf r}_{\mathrm{max}}|)|/|\psi({\bf r}_{\mathrm{max}})|$ for the center of the band, where ${\bf r}_{\mathrm{max}}$ is the location in real space where $|\psi({\bf r})|$ is maximal. 
 We find no real qualitative change in the wave function across the DSM-CDM QCP, in complete agreement with the fact that this is not a localization transition in the wave function. In contrast, we find a drastic change in the qualitative nature of the wave function across the CDM-AI localization transition, where for $W>W_l$, $\Psi({\bf r}) \sim \exp(-|{\bf r} - {\bf r}_{\mathrm{max}}|/\lambda)$ with a finite localization length $\lambda$ in the AI phase.

\begin{figure}[htb]
\begin{minipage}{.4\textwidth}
\centering
\includegraphics[width=0.6\linewidth,angle=-90]{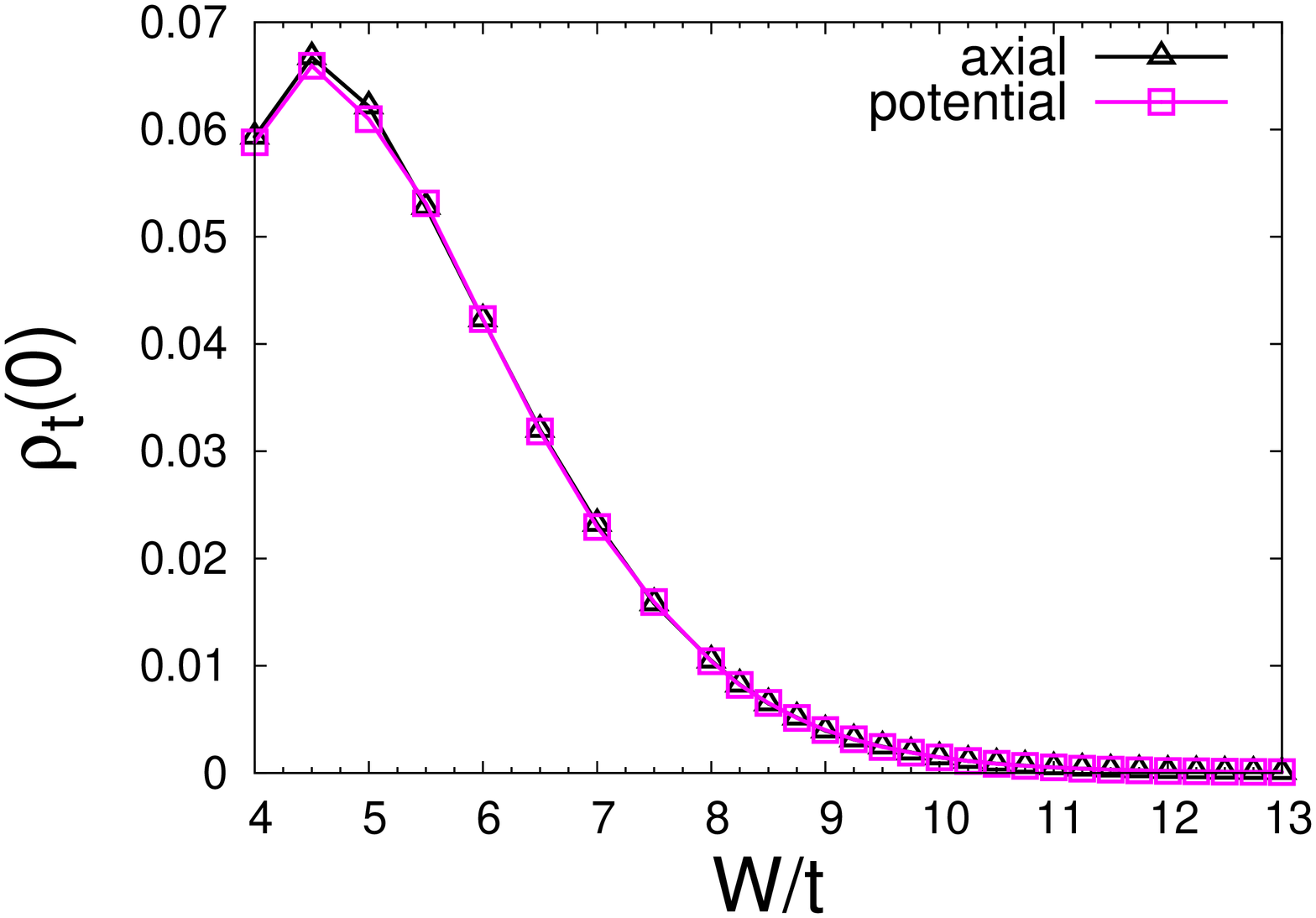}
\end{minipage}
\begin{minipage}{.4\textwidth}
\centering
\includegraphics[width=0.6\linewidth,angle=-90]{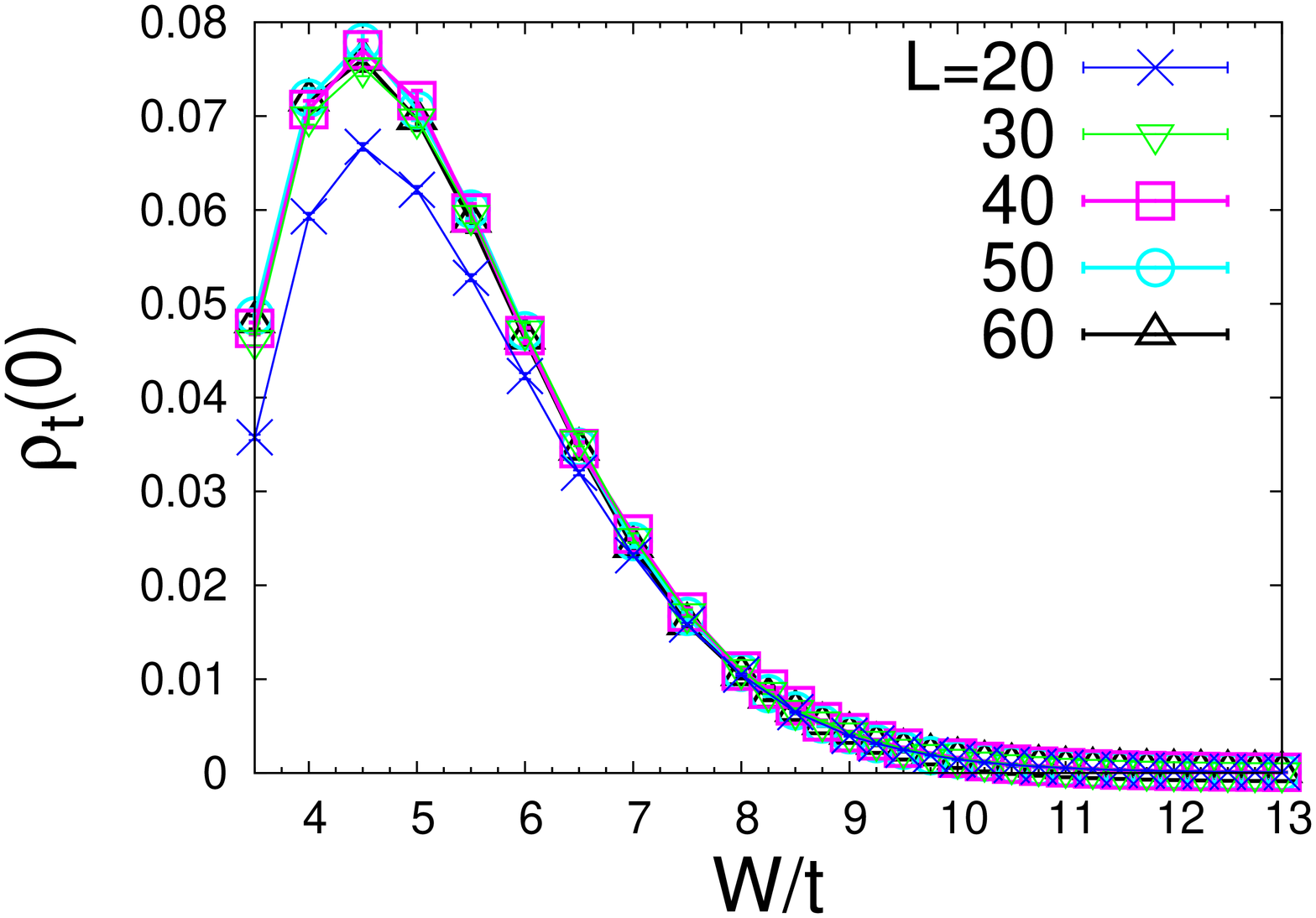}
\end{minipage}
\caption{(Left) The typical density of states at zero energy as a function of disorder $W$ with $L=20$ for axial and potential disorder.  Remarkably, we find the results are equivalent to within numerical accuracy and therefore conclude the critical exponent $\beta$ is the same for these two types of disorder. (Right) Typical density of states for various system sizes, which show a very weak dependence on the system size for $L\gtrsim30$.}
\end{figure}

\begin{figure}[htb]
\begin{minipage}{.4\textwidth}
\centering
\includegraphics[width=0.9\linewidth]{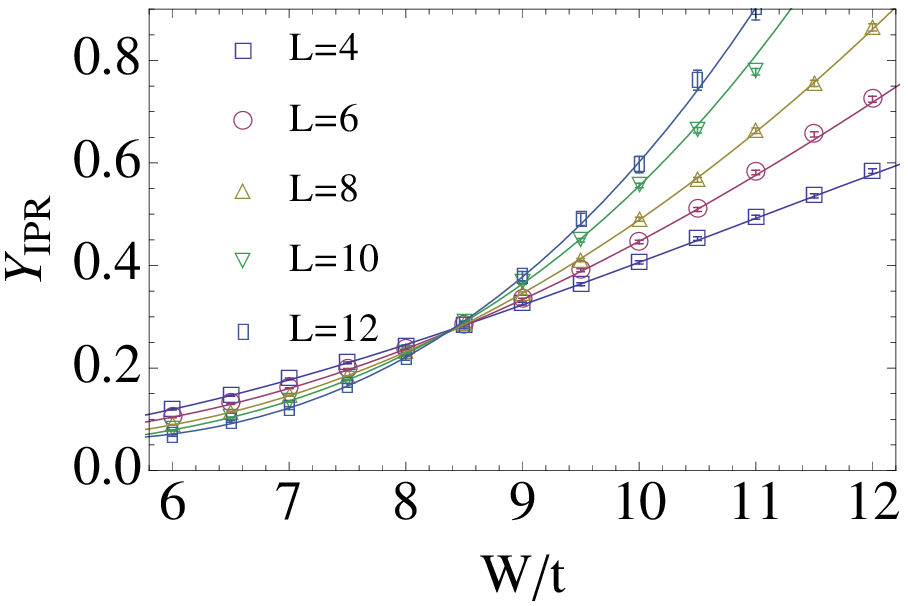}
\end{minipage}
\begin{minipage}{.4\textwidth}
\centering
\includegraphics[width=0.9\linewidth]{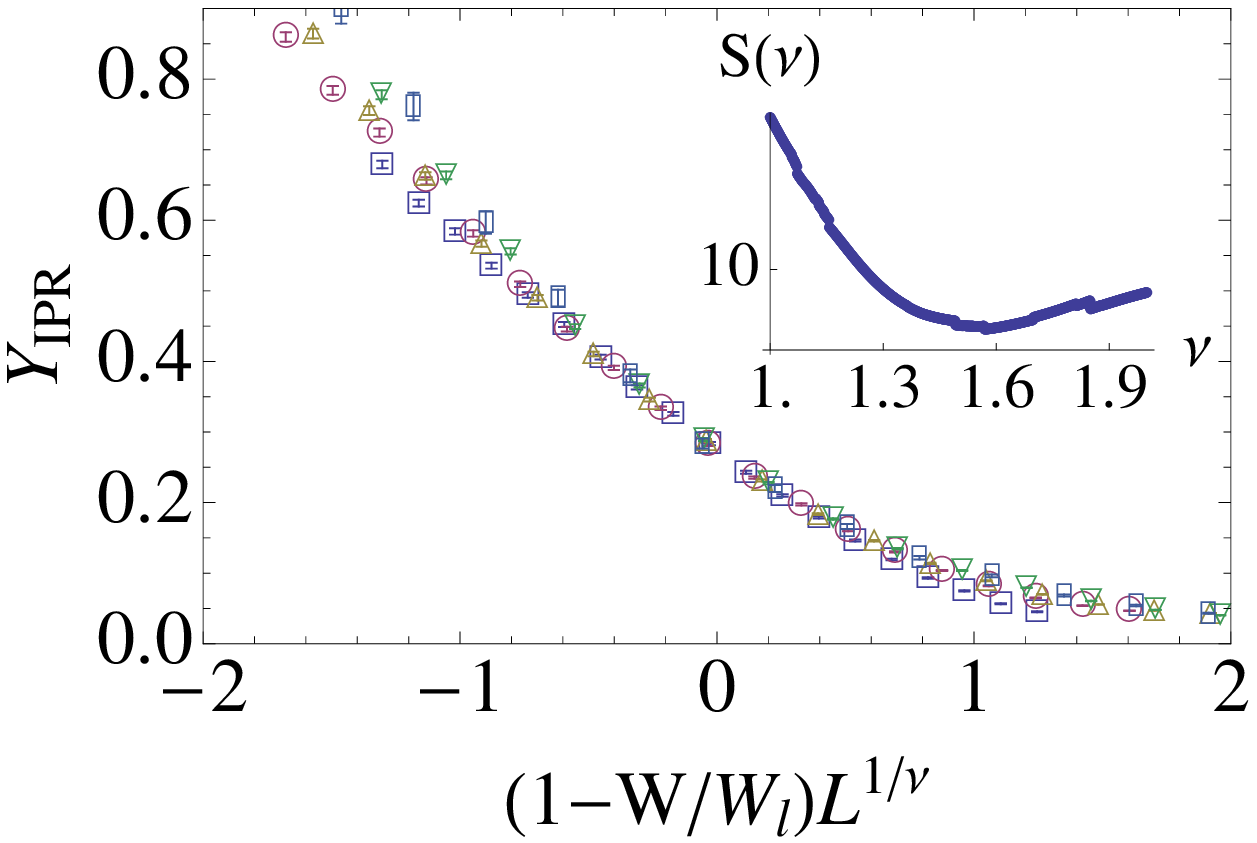}
\end{minipage}
\caption{Finite size scaling of the IPR for potential disorder. (Left) Determining the location of the Anderson localization using third order polynomial fits to obtain $W_l=8.4\pm0.3$, which agrees with the TDOS within numerical accuracy.  We attribute the small finite shift from the TDOS result for $W_l$ to finite size effects.  From this we obtain the critical exponents $d_2 = 1.2 \pm0.3$, $y =2.0\pm0.5$, $A_0=1.8\pm0.6$. (Right) Data collapse of the IPR, which yields $\nu=1.6\pm 0.2$, in excellent agreement with the axial disorder result. (Inset) Quality of the scaling collapse is dictated by minimizing the local linearity function. }
\end{figure}

\begin{figure}[htb]
\begin{minipage}{.5\textwidth}
  \centering
  \includegraphics[width=0.6\linewidth,angle=-90]{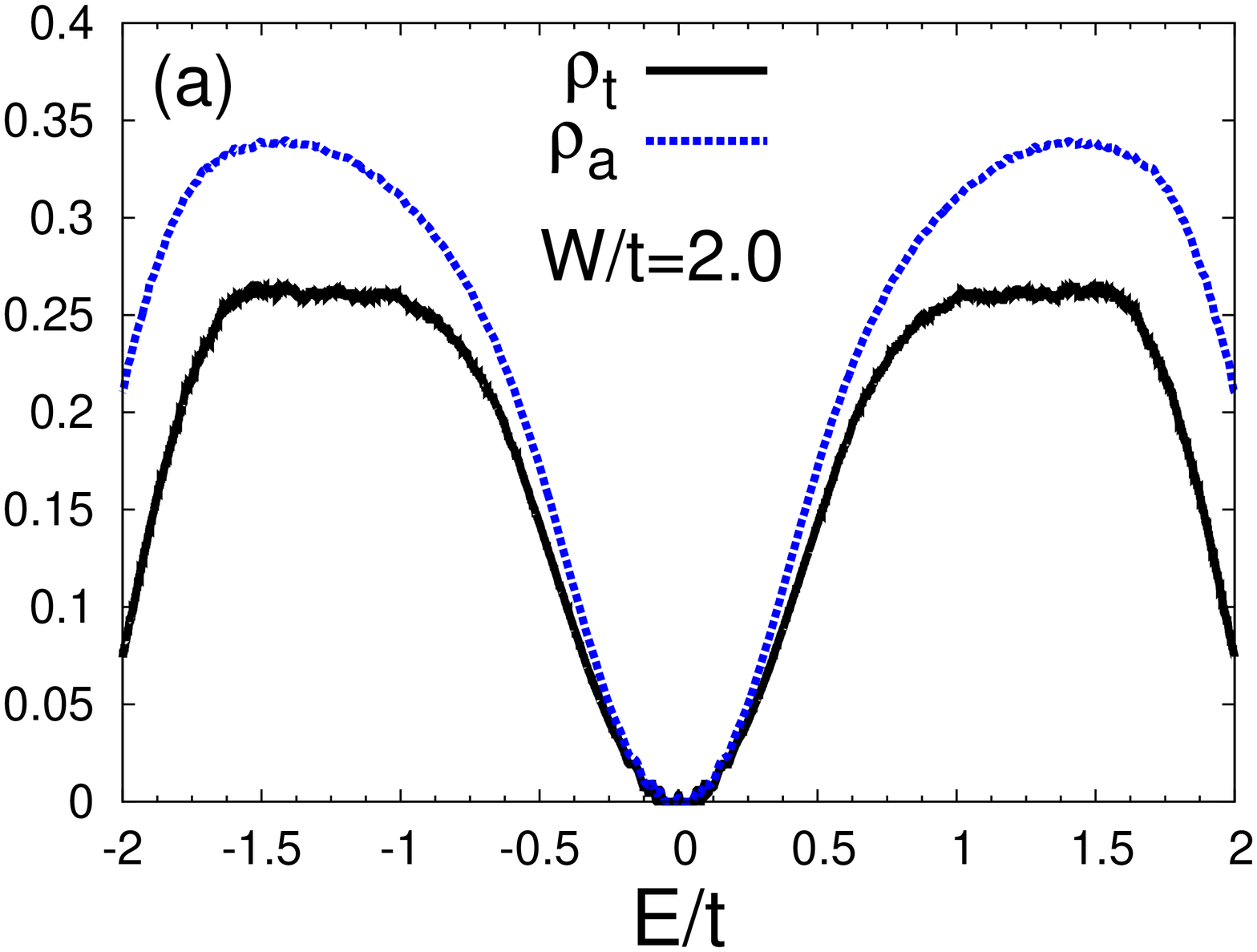}
\end{minipage}%
\begin{minipage}{.5\textwidth}
  \centering
  \includegraphics[width=0.6\linewidth,angle=-90]{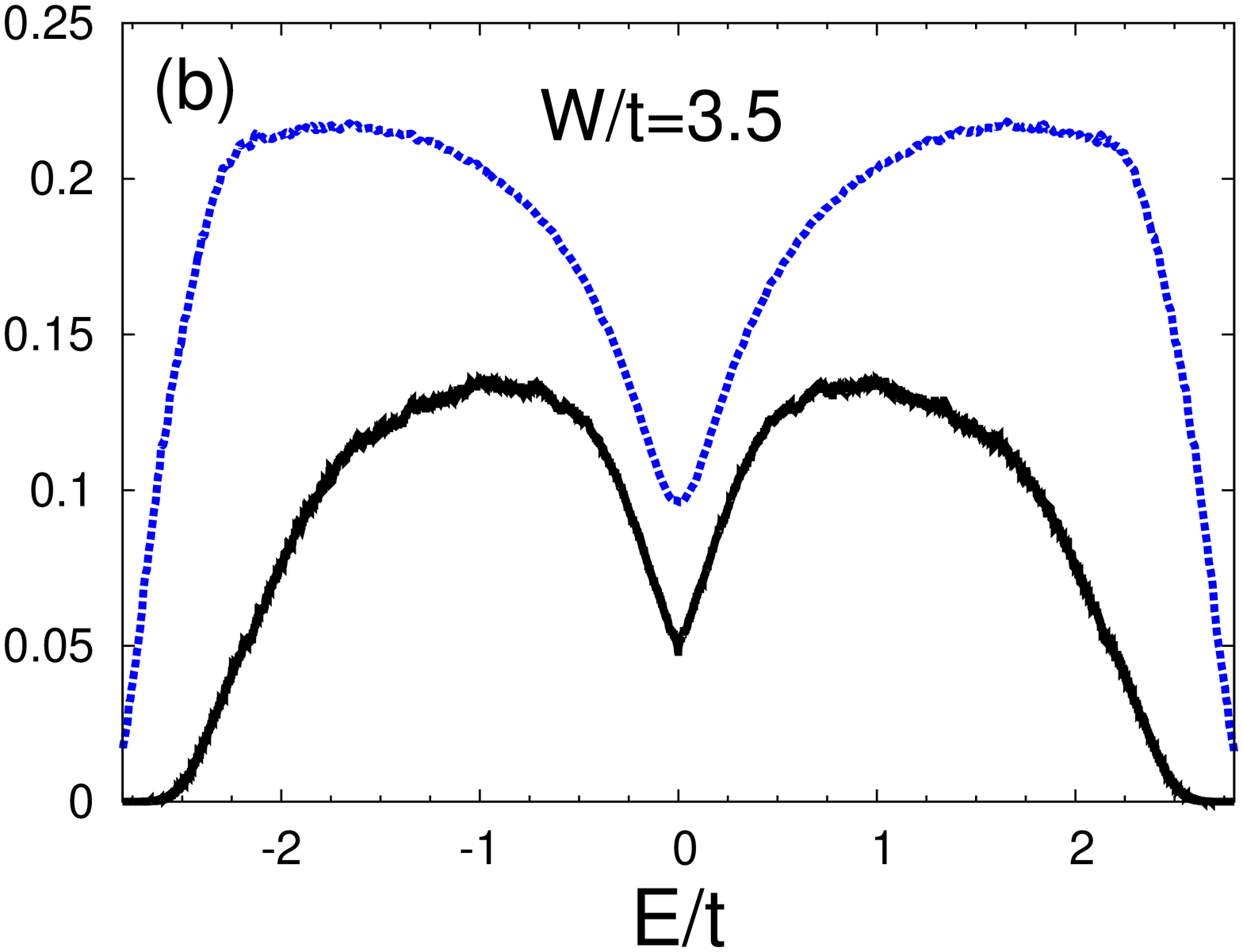}
\end{minipage}
\begin{minipage}{.5\textwidth}
  \centering
  \includegraphics[width=0.6\linewidth,angle=-90]{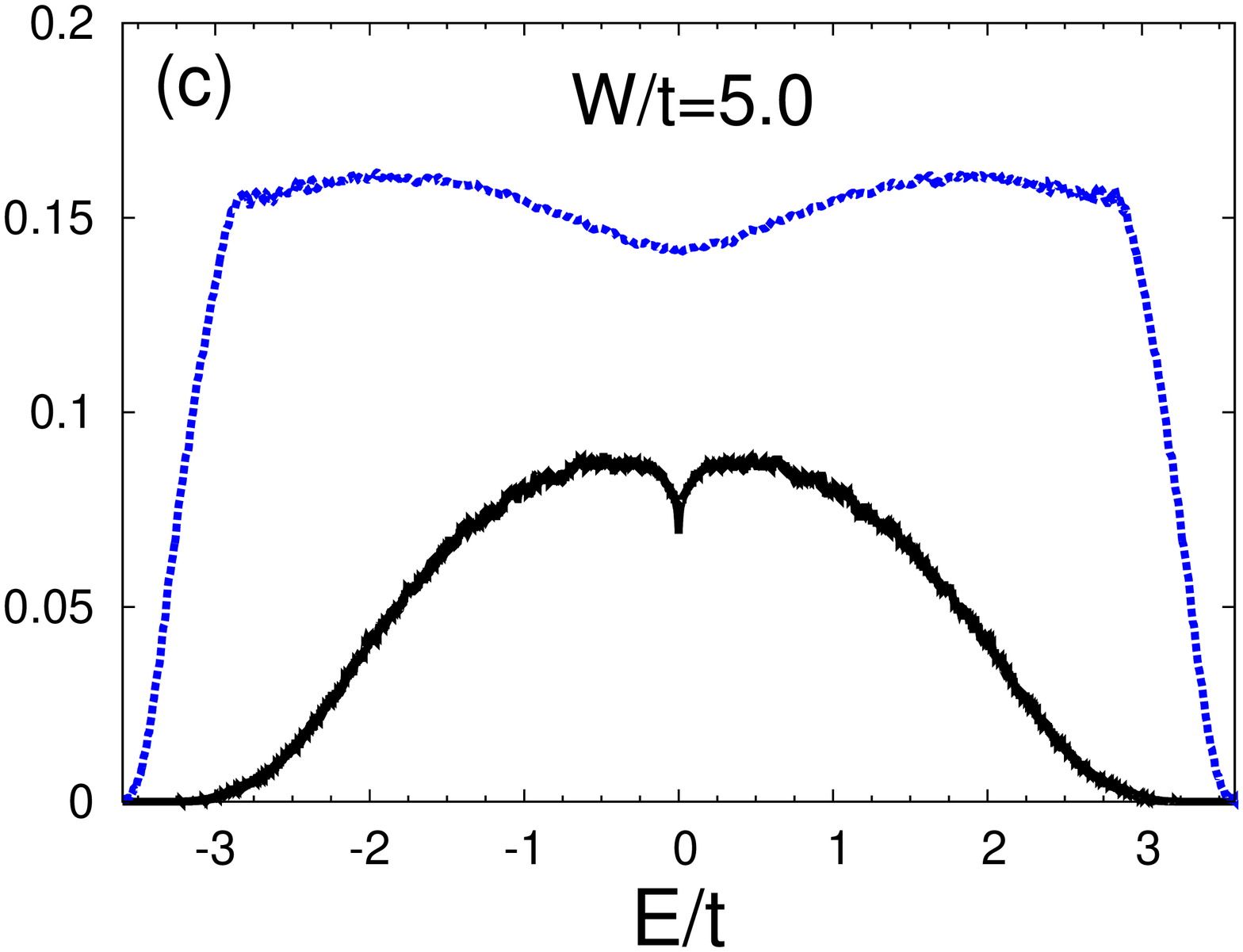}
\end{minipage}%
\begin{minipage}{.5\textwidth}
  \centering
  \includegraphics[width=0.6\linewidth,angle=-90]{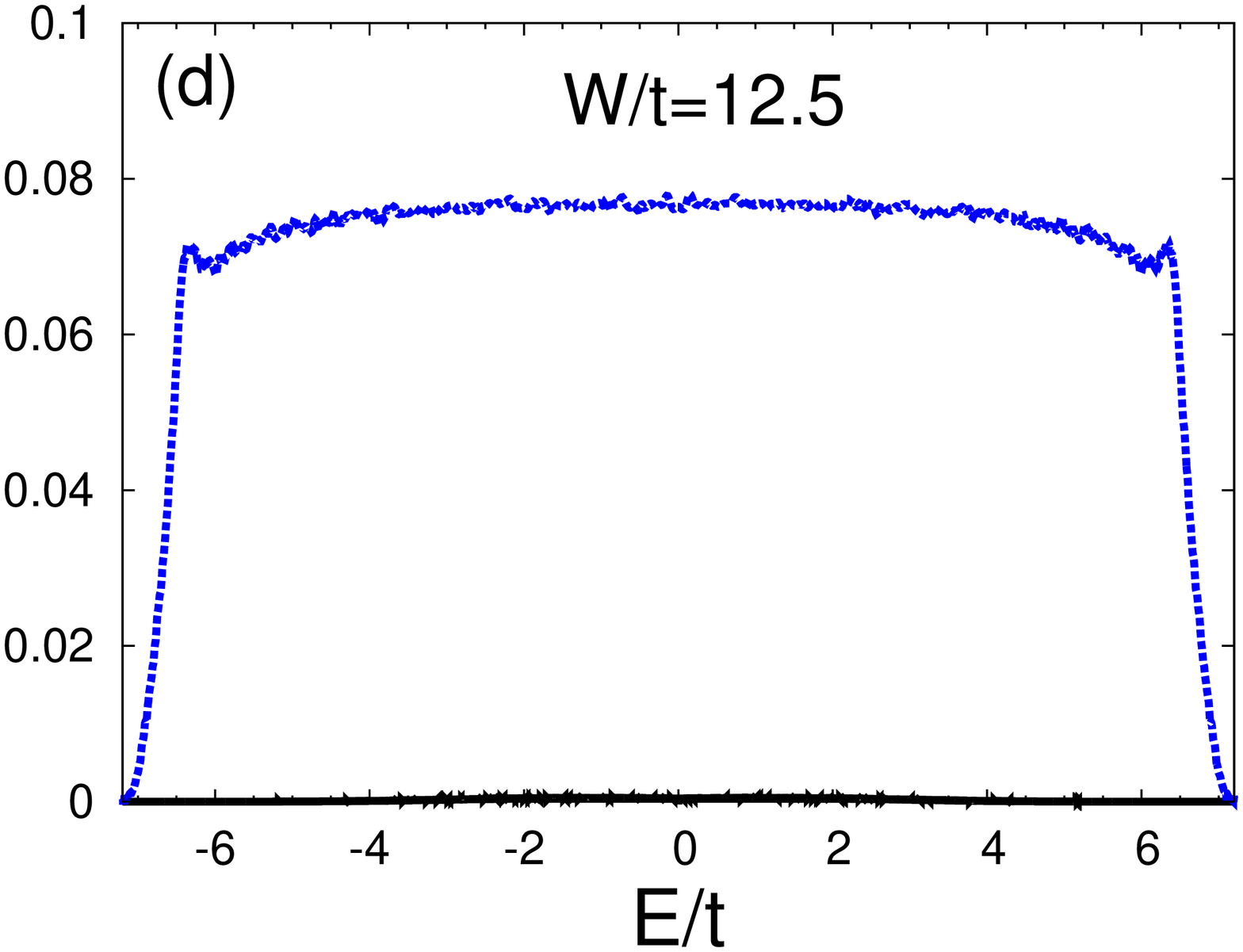}
\end{minipage}
\caption{Comparison of the typical and average density of states with $L=60$ in the Dirac semi-metal phase $W/t=2.0$ (a), the diffusive metal phase $W/t=3.5$ (b) and $W/t=5.0$ (c), and the Anderson insulating phase $W/t=12.5$ (d).  We see the typical density of states tracks the average in both the Dirac phase and diffusive metal phase and goes to zero for all energies in the Anderson insulating phase. }
 \label{fig:dostEall}
\end{figure}

\begin{figure}[htb]
\centering
\includegraphics[width=0.4\linewidth,angle=-90]{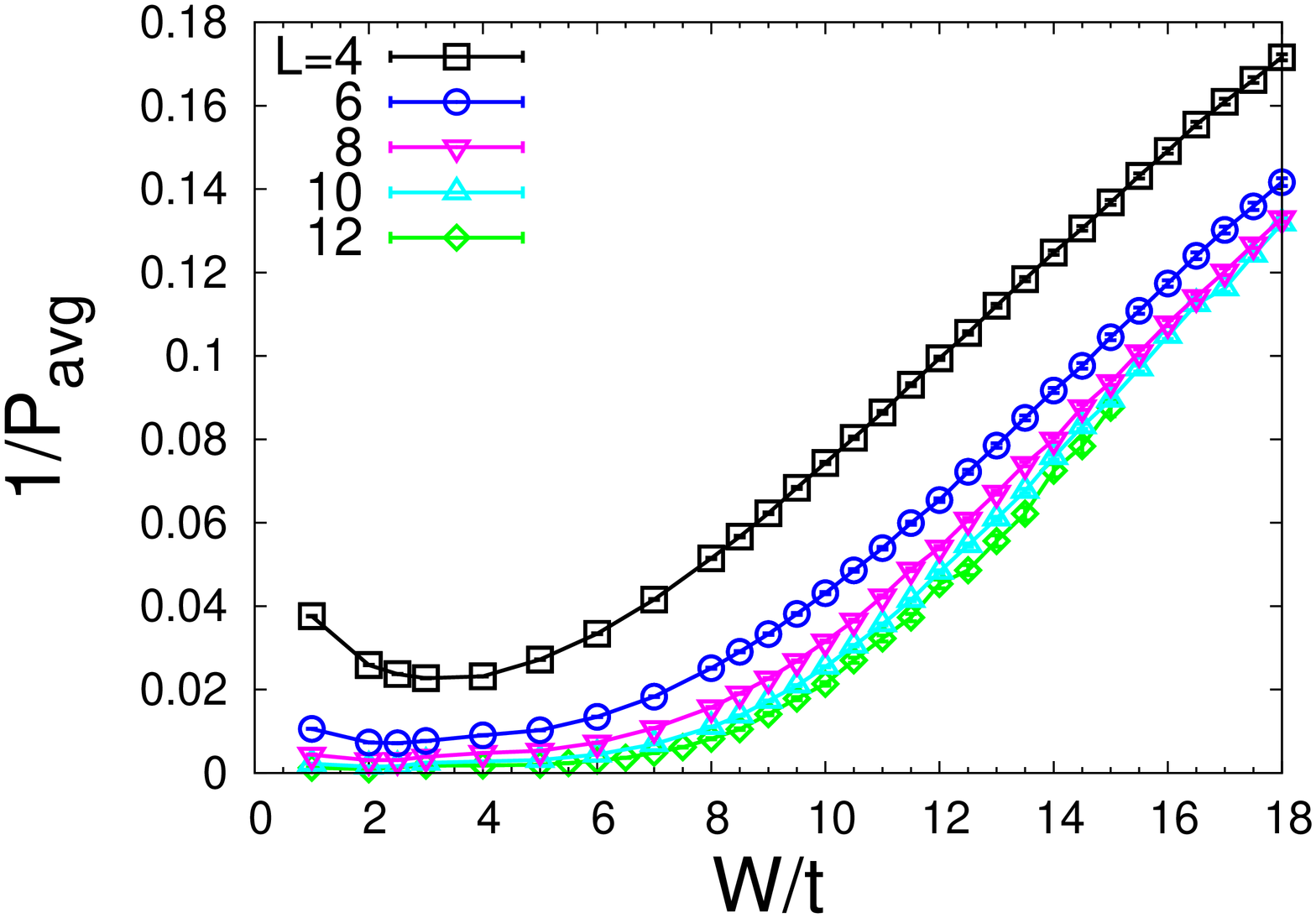}
\caption{Inverse participation ratio versus disorder strength for each system size studied.  We find the expected behavior in a delocalized phase (pertaining to both the DSM and CDM phases) $P_{\mathrm{avg}}^{-1} \sim 1/V$ and in the localized phase approaching a constant.}
\end{figure}

\end{document}